\documentclass{PoS}

\usepackage{subfigure}

\title{Nucleosynthesis of molybdenum in neutrino-driven winds}

\ShortTitle{Nucleosynthesis of Mo in neutrino-driven winds}

\author{\speaker{Julia Bliss}\\
        Institut f\"ur Kernphysik, Technische Universit\"at Darmstadt, Schlossgartenstr. 2,
        64289 Darmstadt, Germany\\
        E-mail: \email{jbliss@theorie.ikp.physik.tu-darmstadt.de}}

\author{{Almudena Arcones}\\
        Institut f\"ur Kernphysik, Technische Universit\"at Darmstadt, Schlossgartenstr. 2,
        64289 Darmstadt, Germany\\
        GSI Helmholtzzentrum f\"ur Schwerionenforschung GmbH, Planckstr. 1, 64291 Darmstadt, Germany \\
        E-mail: \email{almudena.arcones@physik.tu-darmstadt.de}}

\abstract{Neutrino-driven winds that follow core-collapse supernovae are an exciting astrophysical site for the production of heavy elements. Although hydrodynamical simulations show that the conditions
in the wind are not extreme enough for a r-process up to uranium, neutrino-driven winds may be the astrophysical site where lighter heavy elements between Sr an Ag are produced, either by the weak r-process
or by the $\nu \!p$-process. However, it is still not clear if the conditions in the wind are slightly neutron-rich or proton-rich. Therefore, we investigate the nucleosynthesis in the wind for neutron- and
proton-rich conditions and systematically explore the impact of wind parameters on abundances. Here we focus on
molybdenum that has raised attention because several astrophysical scenarios failed to reproduce the solar system (SoS) abundance ratio of $^{92}\mathrm{Mo}$ and $^{94}\mathrm{Mo}$. Moreover, available data of SiC X grains
exhibit different isotopic ratios of $^{95}\mathrm{Mo}$ and $^{97}\mathrm{Mo}$ than in the SoS.
We have investigated if neutrino-driven winds can reproduce the SoS $Y(^{92}\mathrm{Mo})/Y(^{94}\mathrm{Mo})$ and can explain the origin of the $Y(^{95}\mathrm{Mo})/Y(^{97}\mathrm{Mo})$ found in SiC X. 
\
}

\FullConference{XIII Nuclei in the Cosmos,\\
		7-11 July, 2014\\
		Debrecen, Hungary }

\begin{document}

\section{Introduction}

Elements heavier than iron are synthesized by neutron capture reactions due to the increasing Coulomb barrier. There exist two types of neutron capture reactions: slow neutron capture process (s-process) and rapid
neutron capture process (r-process). For a long time it has been thought that the neutrino-driven wind is the astrophysical site for the r-process \cite{Meyer:1992,Woosley.etal:1994}. Today, it is clear that the conditions in the wind are not extreme
enough for a full r-process up to uranium \cite{arcones.janka.scheck:2007,Huedepohl:2009wh, Fischer:2009af}. Nonetheless, neutrino-driven winds are an exciting astrophysical site that can explain the origin
of the lighter heavy elements between Sr and Ag which cannot be understood alone by the r-process (see e.g.~\cite{Qian:2007vq,Travaglio:2003qq}). 
\\
The nucleosynthesis begins at high temperatures ($T \sim 10~\mathrm{GK}$) which keep matter in nuclear 
statistical equilibrium (NSE). There is a balance between nuclear reactions producing seed nuclei and photodissociation reactions destroying those nuclei into nucleons: $(Z, A) \leftrightarrow (A-Z)n +Zp$.
When the expansion continues, slower reactions fall out of equilibrium. At the breakdown of NSE ($T \sim 8-5~\mathrm{GK}$) alpha particles dominate the composition. These alpha particles form seed nuclei by:
$3\alpha \rightarrow  ^{12} \! \mathrm{C}$ or $^{4}\mathrm{He}(\alpha n,\gamma)^{9}\mathrm{Be}(\alpha,n)^{12}\mathrm{C}$. The following alpha process \cite{Takahashi:1994yz,Woosley:1992ek} includes a sequence of 
$(\alpha,\gamma)$,\ $(\alpha,n)$,\ $(\alpha,p)$~reactions in combination with $(n,\gamma)$ and $(p,\gamma)$ reactions. The abundances of neutrons, protons and
seed nuclei depend on the wind parameters. The wind parameters are: entropy, expansion timescale and electron fraction ($Y_{e}$) (see e.g., \cite{Meyer:1993,Qian:1996xt,Hoffman:1996aj,Thompson:2001ys}). 
In case of a neutron-rich wind ($Y_{e} < 0.5$) the weak r-process \cite{Truran:2001zx} occurs, whereas for
proton-rich conditions ($Y_{e} > 0.5$) the $\nu \!p$-process \cite{Pruet:2005qd,Frohlich:2005ys,Wanajo:2006ec} produces heavy elements.
\\
Neutrino-driven winds are also interesting to explain molybdenum.
Among the lighter heavy elements, Mo has raised interest since various astrophysical scenarios failed to reproduce the solar abundance ratio of
$^{92}\mathrm{Mo}$ and $^{94}\mathrm{Mo}$. Furthermore, available data of SiC X grains present different isotopic ratios than those in the solar system.
Molybdenum has seven stable isotopes: the two p-only nuclides $^{92}\mathrm{Mo}$ and $^{94}\mathrm{Mo}$, the mixed s-,r-nuclides $^{95}\mathrm{Mo}$ and $^{97}\mathrm{Mo}$, the s-only nuclide $^{96}\mathrm{Mo}$ and the r-only nuclide $^{100}\mathrm{Mo}$.
$^{92}\mathrm{Mo}$ and $^{94}\mathrm{Mo}$ are shielded from the s- and r-process.
We present here a systematic nucleosynthesis study to identify the necessary conditions to reproduce the observed Mo isotopic ratios based on neutrino-driven winds.

\section{Method}

For our studies we have used a trajectory which corresponds to the explosion model M15l1r (see \cite{arcones.janka.scheck:2007} for more details) and is ejected 5~s after bounce. The M15l1r model has a simple but very efficient neutrino transport.
The initial entropy, expansion timescale and electron fraction of the trajectory are $S=78~\mathrm{k_{B}/nuc}$, $\tau = 5~\mathrm{ms}$ and $Y_{e}=0.484$, respectively. We modify the wind entropy by varing the density (in a radiation dominated environment entropy and density are related by $S \propto T^{3}/\rho$). The expansion timescale is changed
by varying the trajectory time. The calculation of the initial NSE composition requieres an initial electron fraction which is given as a nucleosynthesis parameter. The subsequent evolution of the electron fraction is 
calculated within the network.
Hoffman et al. \cite{Hoffman:1996aj} have shown that the composition resulting from the alpha process only depends on $S^{3}/\tau$ for a given initial electron fraction. Thus, in neutron-rich winds the variation of only one of the two
wind parameters entropy or timescale is sufficient.
\\
The network calculations start at $T \sim 10~\mathrm{GK}$, assuming a NSE composition for the given initial electron fraction. The following evolution of the composition is calculated within a
full reaction network \cite{Frohlich:2005ys} which includes 4053 nuclei from H to Hf including both neutron- and proton-rich isotopes \cite{Arcones:2010yt}. Experimental reaction rates \cite{Angulo:1999} are included when available. Otherwise
theoretical reaction rates from the statistical NON-SMOKER \cite{Rauscher:2000fx} code are used. The theoretical weak reaction rates are the same as in \cite{Frohlich:2005ys} and experimental beta-decay rates \cite{NuDat2} are used when available.

\section{Results}

\begin{figure}[bh]
	\centering
	\subfigure{\includegraphics[width=0.37\textwidth]{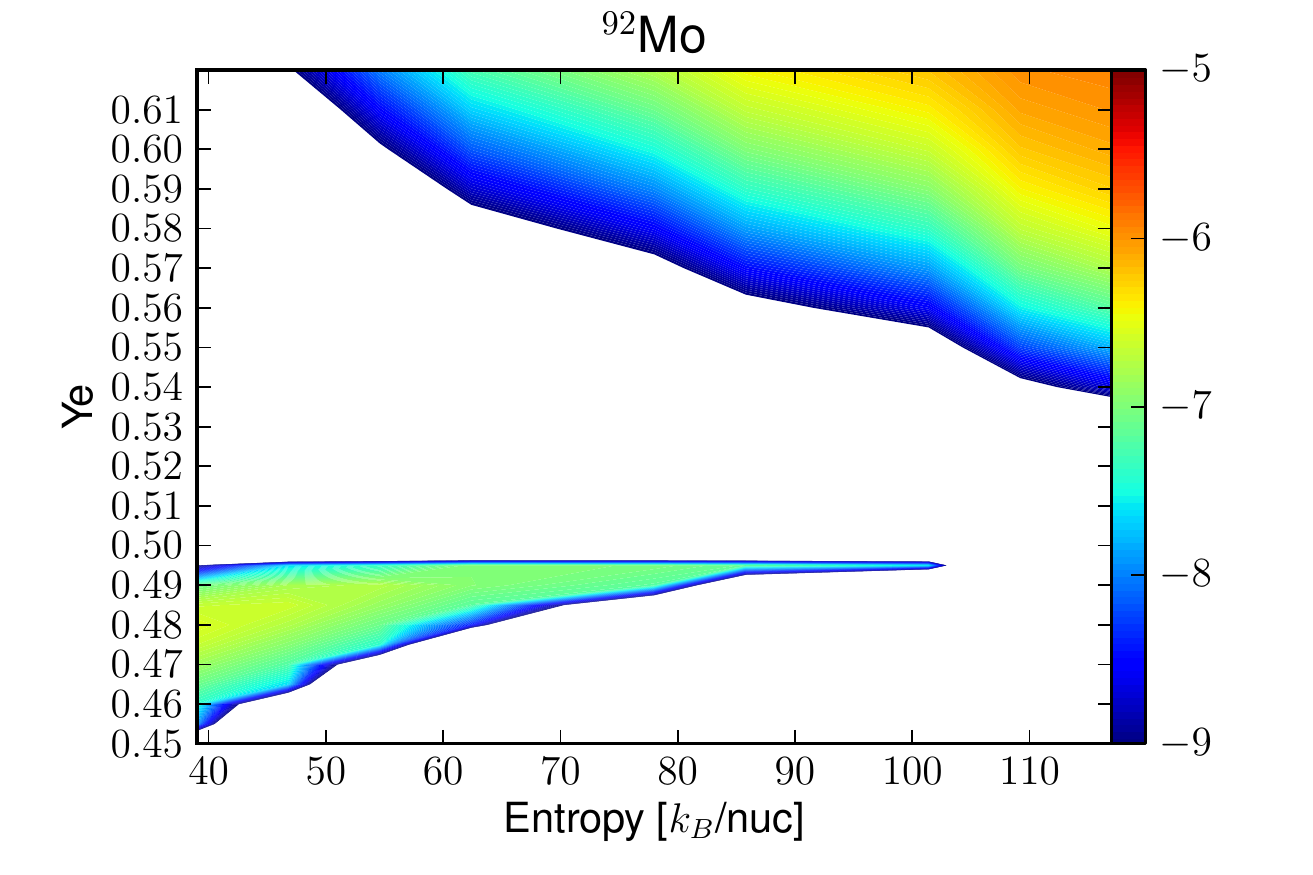}}\quad
	\subfigure{\includegraphics[width=0.37\textwidth]{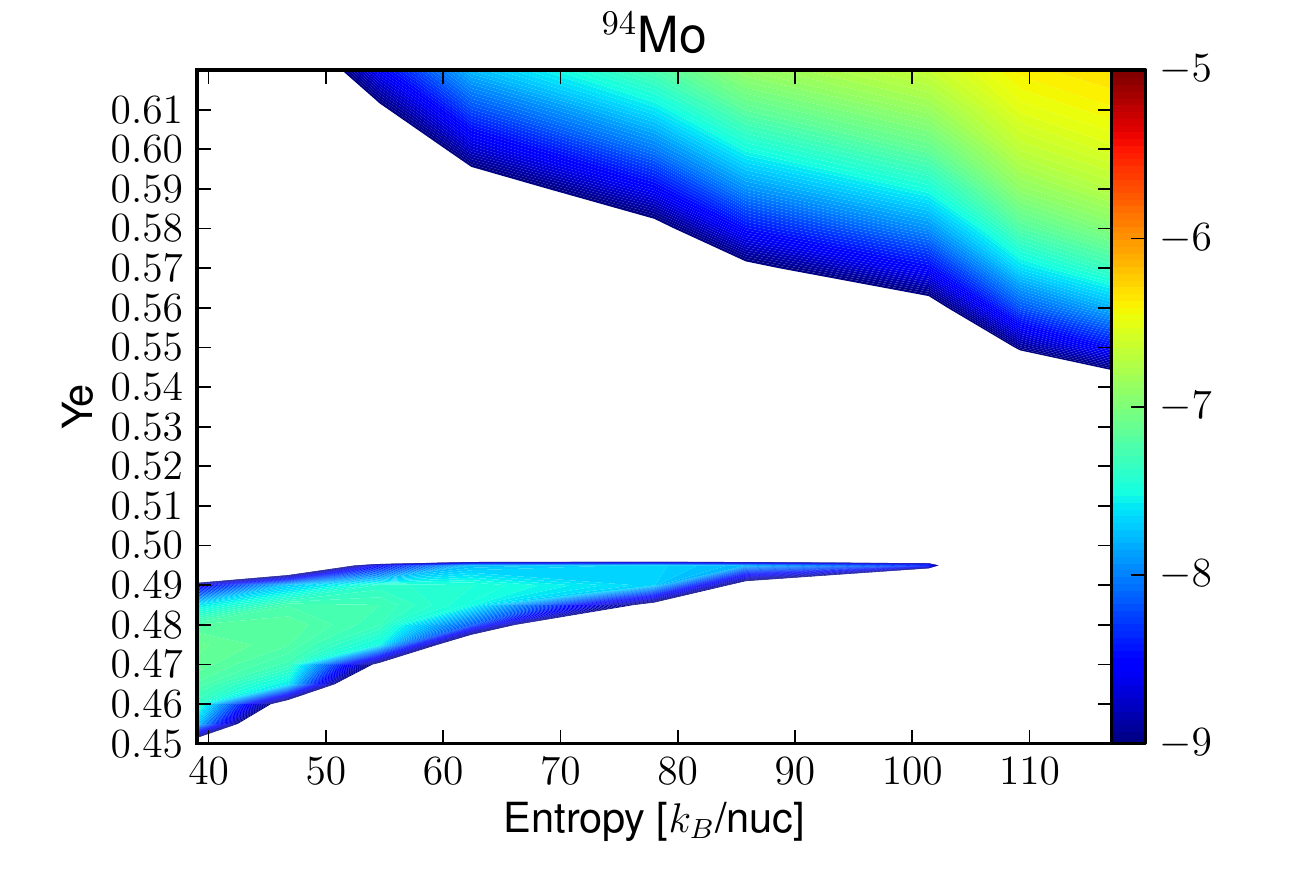}}
	\caption{Colour contours show the abundance of $^{92}\mathrm{Mo}$ (left panel) and $^{94}\mathrm{Mo}$ (right panel) in log scale for different $Y_{e}$ and entropy.}
	\label{Fig:Overview of Mo p-only}
  \end{figure}

\begin{figure}[h]
	\centering
	\includegraphics[width=0.37\textwidth]{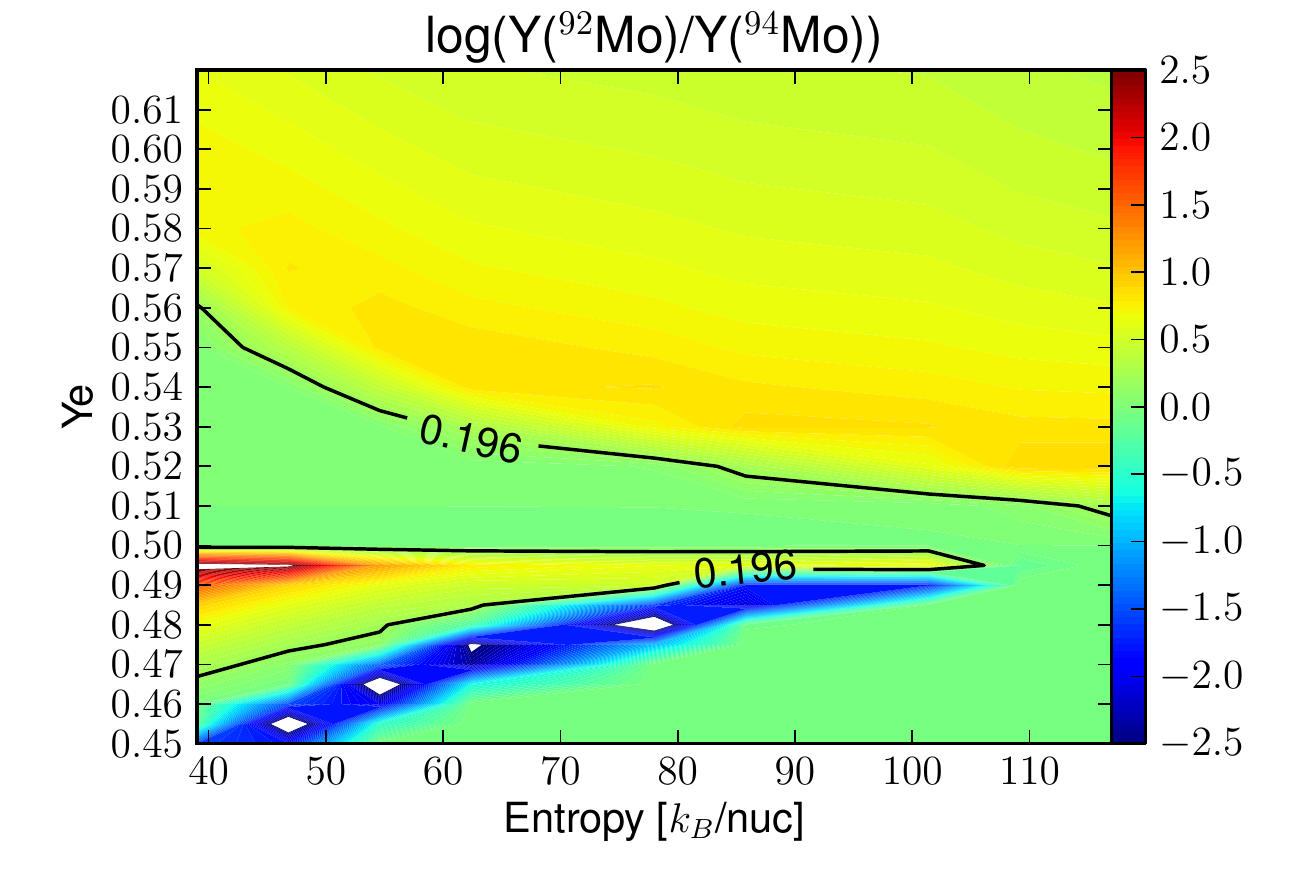}
	\caption{Abundance ratio of $^{92}\mathrm{Mo}$ and $^{94}\mathrm{Mo}$ in log scale for different $Y_{e}$ and entropy. The black line corresponds to the solar ratio: $Y(^{92}\mathrm{Mo})/Y(^{94}\mathrm{Mo})$=1.57.}
	\label{Fig:Overview of ratio 92Mo/94Mo}
  \end{figure}

\begin{figure}[t]
	\centering
	\includegraphics[width=0.37\textwidth]{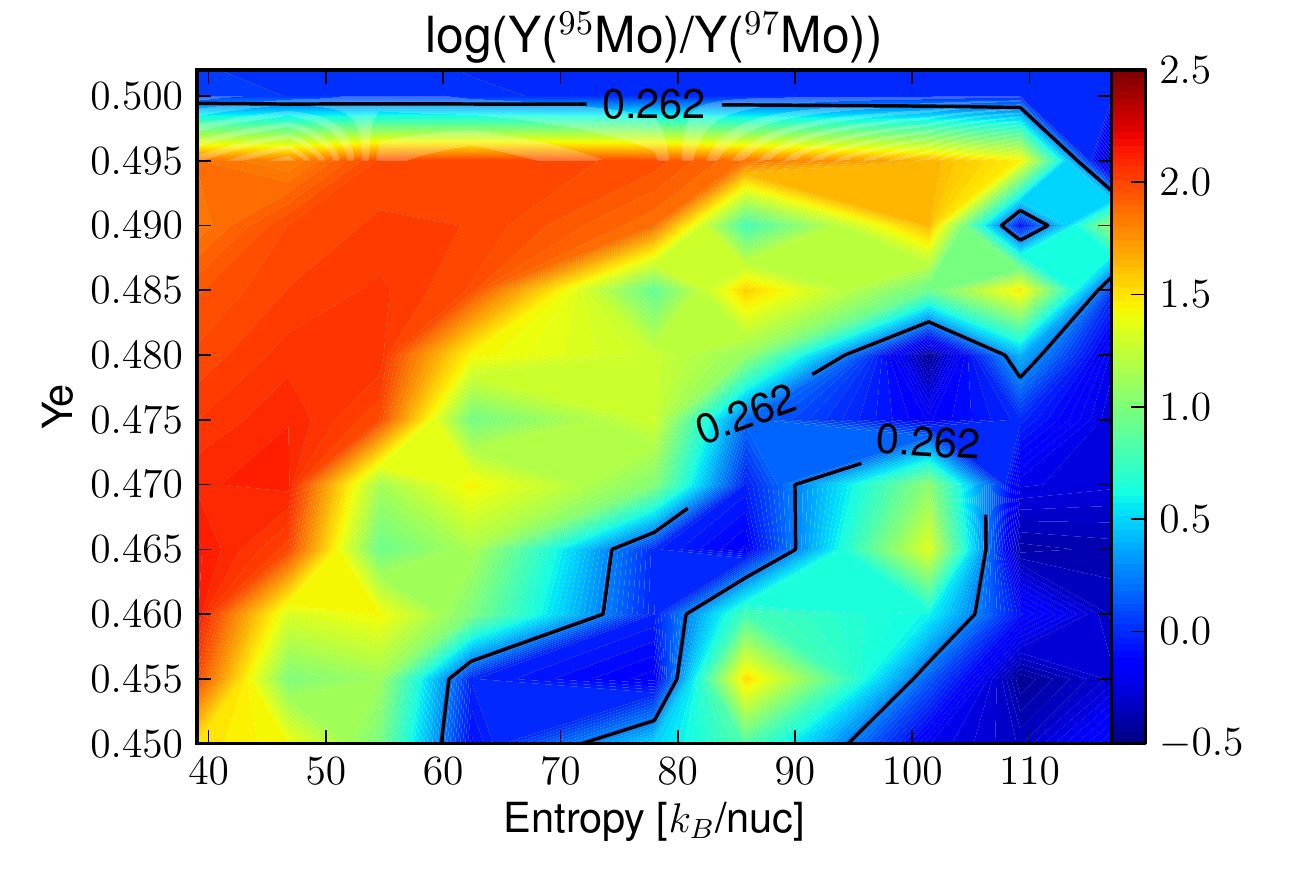}
	\caption{Abundance ratio of $^{95}\mathrm{Mo}$ and $^{97}\mathrm{Mo}$ in log scale depending on $Y_{e}$ and entropy. The black line shows the solution where 
	$Y(^{95}\mathrm{Mo})/Y(^{97}\mathrm{Mo})$=1.83.}
	\label{Fig:Overview of ratio 95Mo/97Mo in SiC}
  \end{figure}

We have performed a systematic nucleosynthesis study to identify the necessary conditions for the synthesis of the different Mo isotopes in neutrino-driven winds. Figs.~\ref{Fig:Overview of Mo p-only}, \ref{Fig:Overview of Mo s-,r-} show the results for $^{92,94}\mathrm{Mo}$,
and $^{95,97}\mathrm{Mo}$, respectively. The color contours correspond the abundances in log scale for different entropies and electron fractions ($Y_{e}$). 
\\
Both p-nuclides, $^{92}\mathrm{Mo}$ and $^{94}\mathrm{Mo}$
can be synthesized in slightly neutron-rich winds as well as in proton-rich winds for a range of different entropies. In general, the abundances of $^{92}\mathrm{Mo}$ are slightly larger than the ones of $^{94}\mathrm{Mo}$
for the same conditions. In Fig.~\ref{Fig:Overview of ratio 92Mo/94Mo} the color contours show $Y(^{92}\mathrm{Mo})/Y(^{94}\mathrm{Mo})$ in log scale for different entropies and and electron fractions. The black line indicates the SoS ratio:
$Y(^{92}\mathrm{Mo})/Y(^{94}\mathrm{Mo})=1.57$ \cite{Lodders:2003}. The SoS ratio can be reproduced based on both proton-rich and slightly neutron-rich winds. In proton-rich winds the SoS ratio can be achieved
for several combinations of the wind entropy and the electron fraction. Nevertheless, when we consider the abundances of $^{92}\mathrm{Mo}$ and $^{94}\mathrm{Mo}$ within these conditions (Fig.~\ref{Fig:Overview of Mo p-only}),
we see that the abundances of both are negligible. Therefore, we conclude that the solar ratio cannot be due to proton-rich winds. This result agrees with Ref.~\cite{Fisker:2009}.
\\
In slightly neutron-rich winds the solar ratio can be also achieved as shown in Fig.~\ref{Fig:Overview of ratio 92Mo/94Mo}. But the consideration of the abundances of $^{92}\mathrm{Mo}$ and $^{94}\mathrm{Mo}$
within these conditions shows that the ratio can only be reproduced for a reduced region of the paramized space. Our results in
slightly neutron-rich winds agree very well with Ref.~\cite{Farouqi:2010ss}.
\\
\\
The s-, r-nuclides $^{95}\mathrm{Mo}$ and $^{97}\mathrm{Mo}$ can be formed within slightly neutron-rich conditions and their abundances grow with increasing neutron-to-seed ratio (i.e. lower $Y_{e}$ and higher entropy). 
Pellin et al. \cite{Pellin:1999} measured the isotopic composition of Mo in silicon carbide particles of type X (called SiC X). SiC X are presolar micron-sized grains that are recovered
from meteorites \cite{Amari:1997, Clayton:1997, Meyer:2000}. They are characterized by isotopic deficiencies of Si and C and by excess in $^{15}\mathrm{N}$ \cite{Meyer:2000}. However, the exact location where X grains are produced is still not clear
\cite{Meyer:2000}: they may have condensed within the interiors of presolar type II supernovae \cite{Amari:1997, Clayton:1997, Meyer:2000}, but they can also originate from asymptotic giant branch (AGB) stars \cite{Meyer:2000}. 
Pellin et al. \cite{Pellin:1999} discovered that the isotopic composition of Mo in SiC X is quite anomalous. Thus, the isotopic pattern exhibits a large excess in the abundances of $^{95}\mathrm{Mo}$ and $^{97}\mathrm{Mo}$ 
without similarly large excess in $^{96,98,100}\mathrm{Mo}$. This behaviour differs from that derived from a pure r- or s-process.
\\
Since the abundance ratio of $^{95}\mathrm{Mo}$ and $^{97}\mathrm{Mo}$ in SiC X grains indicates a mixture of SoS molybdenum and an unknown nucleosynthesis component, we have studied whether the SoS ratio of 
$^{95}\mathrm{Mo}$ and $^{97}\mathrm{Mo}$ can be synthesized within neutron-rich winds. $^{95}\mathrm{Mo}$ and $^{97}\mathrm{Mo}$ both exhibit a s-process contribution in the solar system of 40\% and 47\% \cite{Travaglio:2003qq}, respectively,
that cannot be explained by the neutrino-driven wind. Therefore, we subtract the s-process contribution in the SoS ratio of $^{95}\mathrm{Mo}$ and $^{97}\mathrm{Mo}$ ($Y(^{95}\mathrm{Mo})/Y(^{97}\mathrm{Mo})$=1.66 \cite{Lodders:2003}) in our calculations. The 
reduced ratio depends on the SoS abundances of $^{95}\mathrm{Mo}$ and $^{97}\mathrm{Mo}$ (see \cite{Lodders:2003}) and their remaining weak r-process contributions (see \cite{Travaglio:2003qq}). So we obtain a reduced ratio of $Y(^{95}\mathrm{Mo})/Y(^{97}\mathrm{Mo})=1.88$. 
This reduced SoS ratio is only slightly different from the one measured in SiC X ($Y(^{95}\mathrm{Mo})/Y(^{97}\mathrm{Mo})=1.83$ \cite{Pellin:1999}).
\\
Fig.~\ref{Fig:Overview of ratio 95Mo/97Mo in SiC} shows the abundance ratio of $^{95}\mathrm{Mo}$ and $^{97}\mathrm{Mo}$ in log scale depending on $Y_{e}$ and entropy. Here, the black line indicates the ratio found in SiC X. 
This ratio can be achieved for several combinations of wind entropy and electron fraction. The abundances of $^{95}\mathrm{Mo}$ and $^{97}\mathrm{Mo}$ within these conditions are significant except for the 
region close to $Y_{e}=0.5$. Since the ratio in SiC X grains is only slightly different from the reduced solar ratio, the latter can also be reproduced very well in slightly neutron-rich winds. This is favoured 
by the fact that $^{95}\mathrm{Mo}$ and $^{97}\mathrm{Mo}$ can be synthesized very well during neutron-rich conditions. Thus, neutrino-driven winds offer an alternative to the neutron burst model (see \cite{Meyer:2000}) for the explanation
of the nucleosynthesis origin of $^{95}\mathrm{Mo}$ and $^{97}\mathrm{Mo}$ in SiC X grains.

  \begin{figure}[h]
	\centering
	\subfigure[]{\includegraphics[width=0.37\textwidth]{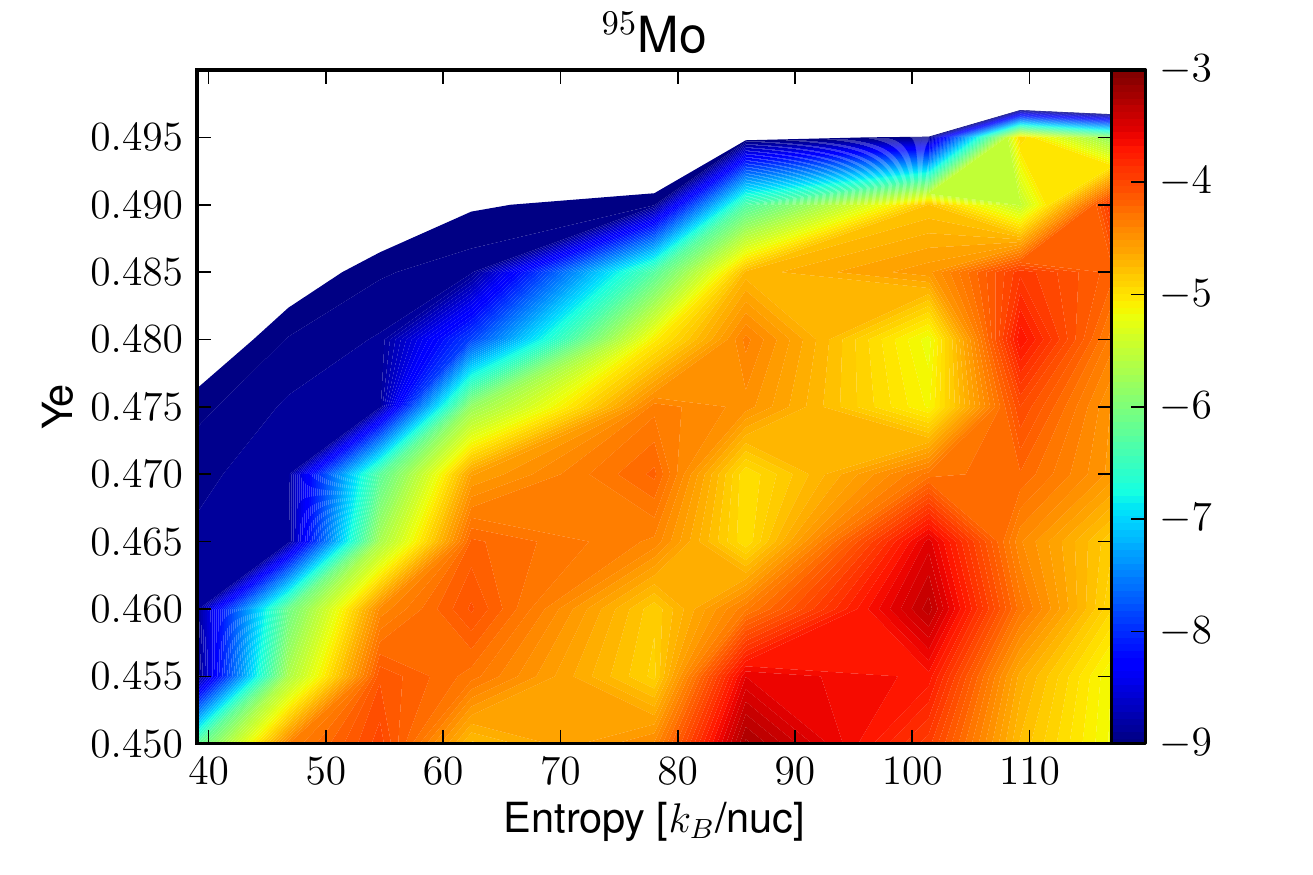}}\quad
	\subfigure[]{\includegraphics[width=0.37\textwidth]{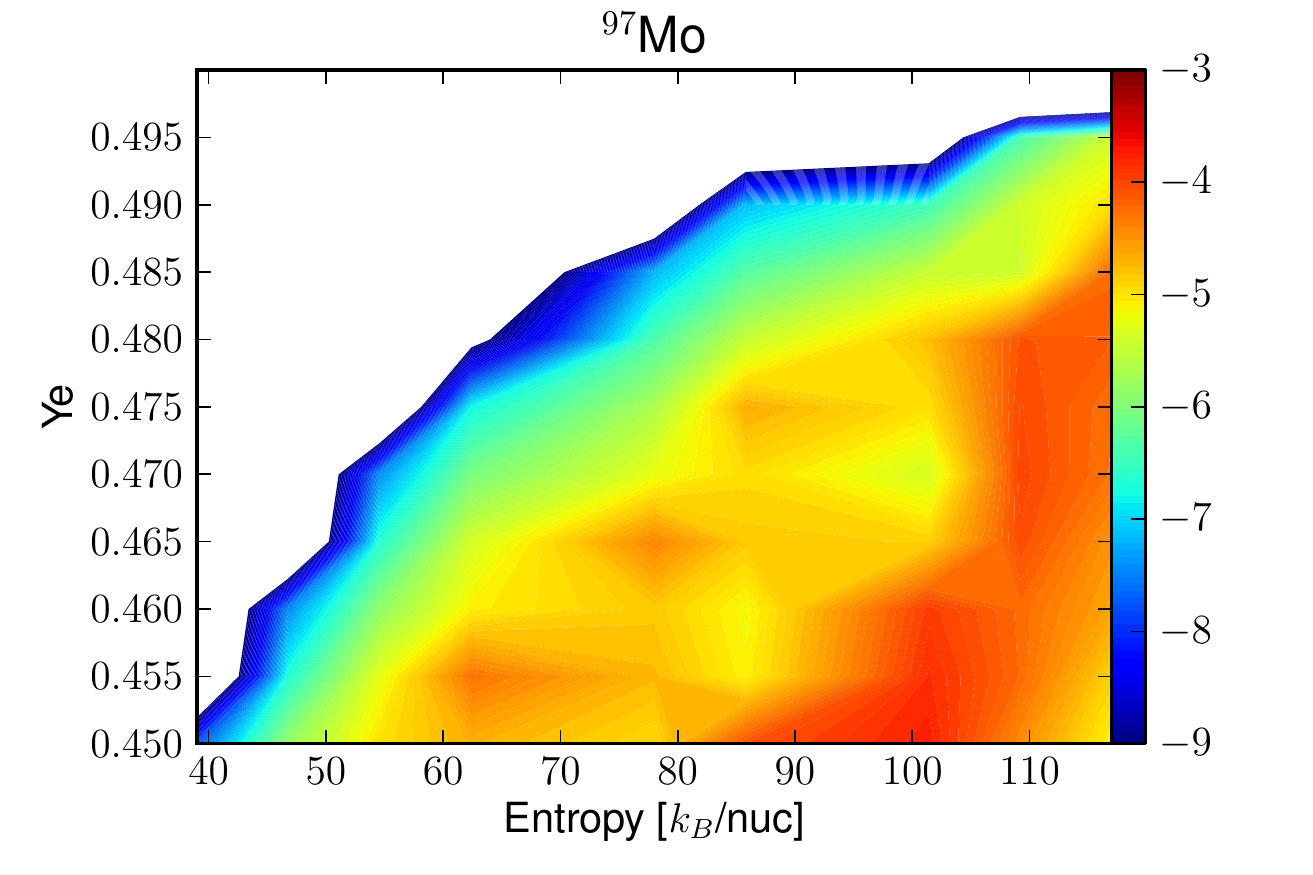}}
	\caption{Colour contours show the abundance of $^{95}\mathrm{Mo}$ (left panel) and $^{97}\mathrm{Mo}$ (right panel) in log scale for different $Y_{e}$ and entropy.}
	\label{Fig:Overview of Mo s-,r-}
  \end{figure}  
  
\section{Summary and Outlook}

Nucleosynthesis studies of neutrino-driven winds combined with observations can give important hints of the origin of elements beyond iron. Even if neutrino-driven winds may not be the production site for 
elements heavier than silver, they may allow for the synthesis of lighter heavy elements in neutron- and proton-rich conditions.
\\
We have done a systematic nucleosynthesis study to identify the required conditions for the formation of molybdenum isotopes based on neutrino-driven winds. $^{92}\mathrm{Mo}$ and $^{94}\mathrm{Mo}$ can be 
formed in slightly neutron-rich as well as in proton-rich conditions. The SoS abundance ratio of the two p-nuclides can only be achieved within neutron-rich conditions for low entropies. In order to conclude 
whether
neutron-rich winds can reproduce the SoS ratio, we will further investigate if the conditions that fulfill the ratio lead to an overproduction of other elements.
\\
For $^{95}\mathrm{Mo}$ and $^{97}\mathrm{Mo}$ we found that they can be synthesized in neutron-rich conditions. Neutrino-driven winds can reproduce the abundance ratio of $^{95}\mathrm{Mo}$ and $^{97}\mathrm{Mo}$ found in SiC X as
well as the SoS abundance ratio (without s-process contributions) of them which is only slightly different from the grain ratio.

\end{document}